\documentstyle[psfig,preprint,aps]{revtex}

\def\<{\langle}
\def\>{\rangle}
\newcommand{\h}{\mathcal{H}}

\newcommand{\non}{\nonumber}
\newcommand{\DS}{\displaystyle}

\newcommand{\bt}{\beta}

\newcommand{\lb}{\lambda}
\def\skip1{\vskip \baselineskip}
\def\skip2{\vskip 2\baselineskip}

\def\beq{\begin{equation}}
\def\eeq{\end{equation}}
\def\beqa{\begin{eqnarray}}
\def\eeqa{\end{eqnarray}}
\begin{document}
\bibliographystyle{prsty1}
\bibliography{refs}

\begin{thebibliography}{99}


\bibitem{by}
Binder K. and Young A.P., {\it Rev. Mod. Phys.,} {\bf 58} (1986) 801.

\bibitem{fh}
Fischer K.H. and Hertz J.A., {\it Spin Glasses} (Cambridge University Press,
Cambridge) 1991.

\bibitem{par}
Parisi G. , {\it Phys. Rev. Lett.,} {\bf 43} (1979) 1754; {\it J. Phys.
A.: Math. Gen.,} {\bf 13} (1980) L115, 1101, 1887.

\bibitem{sk}
Sherrington D. and  Kirkpatrick S., {\it Phys. Rev. Lett.,} {\bf 35} (1975)
1792.

\bibitem{gm}
Gross D.J. and  M\'{e}zard M., {\it Nucl. Phys. B,} {\bf 240} (1984) 431.

\bibitem{der}
Derrida B. , {\it Phys. Rev. Lett.,} {\bf 45} (1980) 79; {\it Phys.
Rev., B} {\bf 24} (1981) 2613.

\bibitem{gar}
Gardner E., {\it Nucl. Phys. B,} {\bf 257} (1985) 747.

\bibitem{tk}
Kirkpatrick T.R. and  Thirumalai D., {\it Phys. Rev. Lett.,} {\bf 58} (1987)
2091; {\it Phys. Rev. B,} {\bf 36} (1987) 5388.

\bibitem{jpb}
Bouchaud J.-P., Cugliandolo L.F., Kurchan J. and M\'{e}zard M.,
in {\it Spin Glasses and Random Fields}, edited by A.P. Young (World
Scientific, Singapore) 1998, pp. 161-223.

\bibitem{hkl}
H\"{o}chli U.T., Knorr K. and Loidl A., {\it Adv. Phys.,} {\bf 39} (1990)
409.

\bibitem{br}
Binder K. and  Reger J.D., {\it Adv. Phys.} {\bf 41}, (1992) 547; Binder K.,
in {\it Spin Glasses and Random Fields}, edited by A.P. Young
(World Scientific, Singapore) 1998, pp. 99-118.

\bibitem{wal}
Walasek K., {\it J. Phys. A: Math. Gen.,} {\bf 28} (1995) L497.

\bibitem{kop}
Kope\'{c} T.K., {\it J. Phys. A: Math. Gen.,} {\bf 29} (1996) L49.

\bibitem{jmed}
de Ara\'ujo J.M. , da Costa F.A. and Nobre F.D., ``Competition Between
Quadrupolar- and Spin-Glass Orderings in Spin-1 Systems'', Preprint, 1999.

\bibitem{mot}
Mottishaw P., {\it Europhys. Lett.,} {\bf 1} (1986) 409.

\bibitem{cal}
Callen H.B., {\it Thermodynamics and an Introduction to Thermostatistics},
Second Edition (John Wiley and Sons, Inc., New York) 1985.


\end{thebibliography}
%Page0 (Title Page)

\newcommand{\bm}[1]{\mbox{\boldmath$#1$}}

\thispagestyle{empty}

%\tightenlines

\author{J.M. de Ara\a'{u}jo,$^1$
F.A. da Costa,$^2$ and F.D. Nobre$^2$ }

\title{First-order transitions and triple point on a random p-spin
interaction model}

\address{
$^1$ Departamento de Ci\^{e}ncias Naturais \\
Universidade Estadual do Rio Grande do Norte,   \\
59610-210 \hspace{8mm} Mossor\'o - RN \hspace{8mm}  Brazil\\
$^2$ Departamento de F\a'{\i}sica Te\a'{o}rica e Experimental \\
Universidade Federal do Rio Grande do Norte \\
Campus Universit\a'{a}rio -- \ C.P. 1641 \\
59072-970 \hspace{8mm} Natal - RN \hspace{8mm} Brazil \\
}

\maketitle

\begin{abstract}
\vskip \baselineskip
The effects of competing quadrupolar- and spin-glass orderings are
investigated on a spin-$1$ Ising model with infinite-range random $p$-spin
interactions. The model is studied through the replica approach and a phase
diagram is obtained in the limit $p\rightarrow\infty$. The phase diagram,
obtained within replica-symmetry breaking, exhibits a very unusual feature
in magnetic models: three first-order transition lines meeting at a commom
triple point, where all phases of the model coexist.

\end{abstract}

\vspace{2cm}

\begin{tabbing}
\=xxxxxxxxxxxxxxxxxx\= \kill
\>{\bf Keywords:} \> Spin glasses;
First-order transitions; \\ \> \> Triple point. \\
\>{\bf PACS Numbers:} \>  05.70.-a, 05.70.Fh, 64.60.-i\\
\end{tabbing}

\newpage

The mean-field theory of Ising spin glasses is quite well
understood at the present \cite{by,fh}. Since the pioneering solution of
Parisi \cite{par} for the infinite-range interaction Ising spin glass, the
so-called Sherrington-Kirkpatrick (SK) model \cite{sk}, a wide variety of
spin-glass systems has been investigated within the
replica approach \cite{by,fh}. In particular, this method was used to solve
a generalization of the SK model with the inclusion of $p$-spin
interactions \cite{gm}, and it was found that the $p\rightarrow \infty$
limit leads to the solution of the random-energy model, introduced
earlier and solved by quite different methods \cite{der}. Since
then, many other $p$-spin interaction models have been studied, motivated
by the fact that they are tractable within mean-field theory, for arbitrary
values of $p$, thus
rendering it possible to analyse both $p \rightarrow 2$ and
$p\rightarrow \infty$ limits \cite{gar}. Another fundamental
aspect of such models is their striking connection to real
structural glasses \cite{tk}. Also, it is usually feasible to study
the dynamical properties of those models, making it possible
to gain some insights in the important ageing phenomena presented by
random systems \cite{tk,jpb}.

An important class of random systems, with many physical realizations, is
that of orientational glasses \cite{hkl,br}. These systems are usually
described in terms of an assembly of discrete spin
variables with quadrupolar random
interactions. Recently, a $p$-spin interaction orientational glass model
was investigated by the replica method \cite{wal}, and it was shown that
Parisi's replica-symmetry-breaking (RSB) scheme could be applied
successfully: indeed, it was found that, in the limit $p \rightarrow \infty$,
the low-temperature behavior of such model may be properly described
through a single-step RSB approach. This conclusion was shown to
be true for another class of models which describe multipolar
glasses \cite{kop}.

Despite all the above-mentioned efforts, much less has been studied on
magnetic models where different kinds of disorder are present. This
represents a
very common situation in physical systems, opening a wide variety of new
problems to be investigated. For spin-1
Ising variables, a simple model including pairwise dipolar and quadrupolar
random interactions led to interesting behavior, with a competition between
quadrupolar- and spin-glass orderings \cite{jmed}.

In the present letter we
investigate a spin-1 Ising model including both dipolar and quadrupolar
random $p$-spin interactions. We
consider an infinite-ranged interacting system, consisting of $N$
spins described through the Hamiltonian,

\beq
{\h} = - \sum_{1\le i_{1} <\cdots<i_{p}\le N}J_{i_{1}i_{2}\cdots
i_{p}}S_{i_{1}} S_{i_{2}}\cdots S_{i_{p}}
- \sum_{1\le i_{1} <\cdots<i_{p}\le N}K_{i_{1}i_{2}\cdots
i_{p}}(S_{i_{1}} S_{i_{2}}\cdots S_{i_{p}})^2,
\eeq

\noindent
where each spin variable can assume the values $0, \pm1$. Both
couplings, $J_{i_{1}\cdots i_{p}}$ and $K_{i_{1}\cdots i_{p}}$ are quenched,
independent and identically distributed random Gaussian variables, with
zero means and variances $J^2p!/(2N^{p-1})$ and $K^2p!/(2N^{p-1})$,
respectively.

It should be mentioned that a spin-1 Ising spin-glass model with $p$-spin
interactions, under a single-ion anisotropy field
$D$ was already studied by Mottishaw \cite{mot}; such a model is identical
to the present one only for $D=K=0$.
Futhermore, the resulting phase diagrams of these two models share
some common features, e.g., they both present three distincts
phases.
However, whereas Mottishaw's model exhibits one first-order and two
continuous critical frontiers, we show that the phase diagram of the
present model displays no continuous lines, being characterized by
three first-order critical frontiers, which meet at a triple point.

Applying the replica method \cite{by,fh} for the model defined through Eq.
(1) and following standard procedures, we get the free-energy density

\beq
-\beta f = \lim_{n\rightarrow0} {1\over n}
G_{n}(q_{ab},\lb_{ab},Q_{ab},\gamma_{ab},R_{a},\xi_{a}),
\eeq

\noindent where

\beqa
G_{n}(q_{ab},\lb_{ab},Q_{ab},\gamma_{ab},R_{a},\xi_{a})
&=&\frac{\bt^2}{4}\sum_{a\neq b}(J^2q_{ab}^p + K^2Q_{ab}^p) +
\frac{\bt^2(J^2+K^2)}{4}\sum_{a}R_{a}^p \non \\
&&- \frac{1}{2}\sum_{a\neq b} (\lb_{ab}q_{ab}+\gamma_{ab}Q_{ab}) -
\sum_{a}\xi_{a}R_{a} \\
&&+\ln {\rm Tr} \exp \left[ \frac{1}{2}\sum_{a\neq b}(\lb_{ab}S^{a}S^{b}
+\gamma_{ab}(S^{a}S^{b})^2) + \sum_{a}R_{a}(S^{a})^2\right] \non ,
\eeqa

\noindent
with $\bt = (k_{B}T)^{-1}$, and $a,b = 1\ldots n$ denoting replica
indices. The quantities $(\lb_{ab},\gamma_{ab},\xi_{a})$ represent
Lagrange multipliers, allowing to fix the set of order parameters
$(q_{ab},Q_{ab},R_{a})$. By demanding $G_n$ to be stationary
with respect to each of those parameters, we get the equilibrium
conditions

\beq
\begin{array}{ccccccccc}
q_{ab} &=& \<S^{a}S^{b}\>, \quad &Q_{ab} &=& \<(S^{a}S^{b})^2\>,
\quad &R_{a} &=& \<(S^{a})^2\>,  \cr
\lb_{ab} &=& \frac{p\bt^2J^2}{2}q_{ab}^{p-1}, \quad
&\gamma_{ab} &=& \frac{p\bt^2K^2}{2}Q_{ab}^{p-1}, \quad
&\xi_{a} &=& \frac{p\bt^2(J^2+K^2)}{4}R_{a}^{p-1}.
\end{array}
\eeq

Throughout most of this letter, we will be interested in the limit
$p \rightarrow \infty$. This is justified by the fact that $p$-interaction
models usually exhibit the same qualitative behavior for finite values of
$p$ \ ($p>2$). Besides that, finite values of $p$ require a substantial
amount of numerical work, whereas in the limit $p \rightarrow \infty$ most
of the calculations may be carried analytically, based on the fact that the
energies of distinct configurations are uncorrelated \cite{gm,der,mot}.

As a preliminary approach to the problem, let us consider the
replica-symmetric (RS) solution, i.e.,
$q_{ab} = q,~~ Q_{ab} = Q,~~ R_{a} = R, ~~\lb_{ab} =
\lb,~~\gamma_{ab} = \gamma, ~~\xi_{a} = \xi,$
in terms of which the equilibrium
equations become

\beq
\lb = \frac{p\bt^2J^2}{2}q^{p-1}, ~~ \gamma =
\frac{p\bt^2K^2}{2}Q^{p-1}, ~~ \xi =  \frac{p\bt^2(J^2+K^2)}{4}R^{p-1},
\label{rs1}
\eeq

\noindent with

\beq
q = \<\varphi_{1}^2\>_{xy} ~ , \quad Q = \<\varphi_{2}^2\>_{xy} ~ , \quad R =
\<\varphi_{2}\>_{xy} ~ .   \label{rs2}
\eeq
In the equations above, $\<(...)\>_{xy}$ stand for
$\int\int_{-\infty}^{+\infty}\frac{dxdy}{2\pi}{\rm exp}(-{x^2+y^2 \over 2})
(...)$,
and

\beq
\varphi_{1} = \frac{2{\rm e}^{\delta}\sinh(\sqrt{\lb}x)}{Z}~, \quad
\varphi_{2} = \frac{2{\rm e}^{\delta}\cosh(\sqrt{\lb}x)}{Z}~,
\eeq
where
\beq
\delta = -\frac{\lb+\gamma}{2}+\xi+\sqrt{\lb}y~, \quad
Z = 1 + 2{\rm e}^{\delta}\cosh(\sqrt{\lb}x)~.
\eeq
One may easily see that the order parameters $Q$ and $R$ never vanish and
that Eqs. (\ref{rs1}) and (\ref{rs2}) always present a trivial solution with
$q=\lambda=0$, for arbitrary values of $p$; herein we identify such a
solution with a quadrupolar-glass (QG) phase \cite{jmed}.

In the limit $p \rightarrow \infty$ the QG solution represents the only
acceptable solution associated with the parameter $q$, since any solution
with $q \ne 0$ is unstable, similarly to what happens in the corresponding
spin-${1 \over 2}$ model \cite{gar}; however, there are two simple
solutions for the parameters $Q$ and $R$, as we describe below.
The first one is given by $\gamma = \xi = 0$, and $Q=R^2=4/9$, in which case
the free-energy density becomes

\beq
f=-k_{B}T \ln3~, \label{sol1}
\eeq
corresponding to an entropy per spin $s = k_{B} \ln 3$; we shall refer to
the phase described by such a solution as a quadrupolar-glass 1 (QG1).
A second solution can easily be found with $Q=R=1$,
$2\gamma = \bt^2K^2p$, and $4\xi = \bt^2(J^2+K^2)p$, in such a way as to yield
a free-energy density,

\beq
f = -\frac{J^2}{4k_{B}T} - k_{B}T\ln 2 ~ .
\eeq
This solution, which we call quadrupolar-glass 2 (QG2), presents
an entropy per spin which becomes negative for
$ k_{B}T/J < k_{B}T_c/J =\frac{1} {2\sqrt{\ln2}}=0.6005\cdots $.
For $T > T_{c}$, both solutions are stable,
the former one (QG1) presenting a lower free energy at
high temperatures. As the temperature is lowered, we find a first-order
transition line, where the free energies of those solutions coincide;
this line is independent of $K$ and is given by
$k_{B}T_{1}/J = \frac{1}{2\sqrt{\ln(3/2)}}= 0.7852\cdots$. It is important
to mention that we have also found other solutions,
all of them being completely unstable. Therefore, in the limit
$p \rightarrow \infty$ the RS solution leads to the phase diagram exhibited
in Fig. 1, with two quadrupolar-glass phases,
QG1 ($q = 0, Q = 4/9, R = 2/3$) and QG2 ($q= 0, Q=R= 1$), separated by a
first-order transition line.

Since the RS Ansatz leads to a QG2 solution which becomes unstable at
low temperatures, one must
carry on with a RSB procedure. In analogy with the spin-${1 \over 2}$
corresponding
problem \cite{gm,gar} and other $p$-interaction orientational glasses
\cite{wal,kop}, one may
see that it is sufficient to consider a single-step Parisi RSB scheme for
the present problem. This is achieved by grouping the
$n$ replicas into $n/m$ blocks of $m$ replicas each. Order parameters with
replica indices
$a,b$ in the same block take on certain values $(q_{ab}=q_{1},~Q_{ab}=Q_{1},
~\lb_{ab}=\lb_{1},~\gamma_{ab}=\gamma_{1})$, distinguished from those
with replica indices in different blocks
$(q_{ab}=q_{0},~Q_{ab}=Q_{0},~\lb_{ab}=\lb_{0},~\gamma_{ab}
=\gamma_{0})$, whereas the single-replica-index parameters are considered
in the RS approximation $(R_{a}=R,~\xi_{a}=\xi)$.
As usual \cite{par}, in the limit $n\rightarrow 0$ the parameter $m$ becomes
a continuous variable in the interval $[0,1]$. For an arbitrary
value of $p$, the free-energy density becomes

\beqa
\bt f = &-{\DS\frac{(\bt J)^2}{4}}\left[(m-1)q_{1}^p-mq_{0}^p\right]
        -{\DS\frac{(\bt K)^2}{4}}\left[(m-1)Q_{1}^p-mQ_{0}^p\right]
        -{\DS\frac{\bt^2(J^2+K^2)}{4}}R^p \non \\
        &+\frac{1}{2}\left[(m-1)(\lb_{1}q_{1}+\gamma_{1}Q_{1})
          -m(\lb_{0}q_{0}+\gamma_{0}Q_{0})\right] + \xi R \non \\
        &+{\DS\frac{1}{m}\int_{-\infty}^{\infty} \int_{-\infty}^{\infty}
        \frac{dx_{0}dx_{1}}{2\pi}}{\rm exp}({-\frac{x_0^2+x_1^2}{2}})
        \ln Z(x_0,x_1) ,
\eeqa
where

\beq
    Z(x_0,x_1) ={\DS\frac{1}{m}\int_{-\infty}^{\infty} \int_{-\infty}^{\infty}
                \frac{dy_{0}dy_{1}}{2\pi}}{\rm exp}({-\frac{y_0^2+y_1^2}{2}})
                [{\rm Tr (e^B)}]^m ,
\eeq
and

\beq
     B = (\sqrt{\lb_0}x_0 +\sqrt{\lb_1-\lb_0}y_0)S +
         (\sqrt{\gamma_0}x_1 +\sqrt{\gamma_1-\gamma_0}y_1 + \xi -
          \frac{\lb_1}{2} - \frac{\gamma_1}{2})S^2.
\eeq
From the above free-energy density functional we can obtain
several solutions, including those already described within the RS
approximation. From now on, we will restrict ourselves to the limit
$p \rightarrow \infty$. Our analysis indicates that there is only one new
solution within a one-step RSB that is physically acceptable, i.e.,
$q_0, Q_0 < 1, q_1 = Q_1 = R = 1$. This solution presents a
a free-energy density independent of $T$, given by

\beq
        f = -\sqrt{(J^2+K^2)\ln2} ~,
\eeq
with a zero entropy. It corresponds to the low-temperature
phase, where all spin variables are frozen completely at random, each of them
in one of the states $S_{i} = \pm 1$, like in a spin-${1 \over 2}$ Ising
spin glass. Such a solution will be associated
with the quadrupolar-spin-glass (QSG) phase.

Since we have obtained three phases (QG1, QG2 and QSG) and their
respective free energies, we can draw the phase diagram of the
model within the RSB approach (see Fig. 2). We adopt the standard
thermodynamic criteria, i.e., whenever two or more solutions are stable,
the correct phase is defined as the one with the
lowest free energy. Besides the first-order transition line
separating phases QG1 and QG2 (which remains a line independent of $K$ at
$k_{B}T_{1}/J=0.7852 \cdots $), we find two new first-order
transition lines: one represents the coexistence of phases QG2 and QSG and
is given by

\beq
       k_{B}T_{2} = {\DS \frac{\sqrt{J^2+K^2}+K}{2\sqrt{\ln2}}} ~,
\eeq
whereas the other one corresponds to a coexistence of phases QG1 and QSG,

\beq
      k_{B}T_{3} = {\DS \frac{\sqrt{(J^2+K^2)\ln2}}{\ln3}}~.
\eeq
The three phases coexist at a triple point, whose coordinates are given by

\beq
     k_{B}T_{t}/J =     \frac{1}{2 \sqrt{\ln(3/2)}}= 0.7852\cdots,
     \quad
     K_{t}/J = {\DS \frac{2\ln2 - \ln 3}{2\sqrt{\ln2 \ln(3/2)}}} =
     0.2713\cdots ~,
\eeq
where all three lines merge together, in accordance
with the famous Gibbs phase rule \cite{cal}.

In conclusion, we have solved a disordered spin-1 Ising model with $p$-spin
interactions through the replica method. Considering the limit
$p \rightarrow \infty$, we have verified that the
Parisi Ansatz is suitable to determine correctly a
phase diagram with genuine first-order transitions, each of them
accompanied by a latent heat and exhibiting discontinuities on the
respective order parameters. Each phase is described by the solution which
is a global minimum of the free energy. A very uncommon feature in magnetic
models has been detected, i.e., a triple point where all three phases
coexist. An important aspect to be
explored is the dynamics of this model, mainly along the
first-order transition lines; this issue is currently been investigated.

\vspace{.5in}
\noindent
{\bf Acknowledgments} \\
One of us (FDN) thanks CNPq and Pronex/MCT (Brazil) for partial financial
support.

%%%%%%%%%%%%%%%%%%%%%%%%%%%%%%%%%%%%%%%%%%%%%%%%%%%%%%%%%%%%%%%%%%%%%
%%%%%%%%%%%%%%%       figures captions        %%%%%%%%%%%%%%%%%%%%%%%
%%%%%%%%%%%%%%%%%%%%%%%%%%%%%%%%%%%%%%%%%%%%%%%%%%%%%%%%%%%%%%%%%%%%%

\begin{figure}

\vskip \baselineskip

%\centerline{\psfig{figure=diag2.ps,height=8cm,width=8cm}}

\caption{\it The phase diagram within the RS approximation. The
quadrupolar-glass phases, {\bf QG1} and {\bf QG2} (defined in the text),
are separated by a
first-order transition line. Throughout the gray region the solution
{\bf QG2} becomes unstable, presenting a negative entropy.}

\label{diag1}

\end{figure}

\vskip \baselineskip

\noindent
\begin{figure}

%\centerline{\psfig{figure=diag2.ps,height=8cm,width=8cm}}

\caption{\it
The phase diagram within Parisi's RSB procedure. All phase
boundaries are first-order transition lines. At the triple point (black dot)
the three phases coexist.}

\label{diag2}

\end{figure}

\end{document}